# A novel deep learning-based method for monochromatic image synthesis from spectral CT using photon-counting detectors

Ao Zheng[1,2], Hongkai Yang[1,2], Li Zhang[1,2] and Yuxiang Xing[1,2]

[1] Department of Engineering Physics, Tsinghua University, Beijing, 100084, China
[2] Key Laboratory of Particle & Radiation Imaging (Tsinghua University), Ministry of Education, Beijing, 100084, China

E-mail: xingyx@mail.tsinghua.edu.cn

**Abstract**

With the growing technology of photon-counting detectors (PCD), spectral CT is a widely concerned topic which has the potential of material differentiation. However, due to some non-ideal factors such as cross talk and pulse pile-up of the detectors, direct reconstruction from detected spectrum without any corrections will get a wrong result. Conventional methods try to model these factors using calibration and make corrections accordingly, but depend on the preciseness of the model. To solve this problem, in this paper, we proposed a novel deep learning-based monochromatic image synthesis method working in sinogram domain. Different from previous deep learning-based methods aimed at this problem, we designed a novel network architecture according to the physical model of cross talk, and it can solve this problem better in an ingenious way. Our method was tested on a cone-beam CT (CBCT) system equipped with a PCD. After using FDK algorithm on the corrected projection, we got quite more accurate results with less noise, which showed the feasibility of monochromatic image synthesis by our method.

## 1. Introduction

CT imaging technology has been playing an important role in clinical diagnosis since its invention in 1970s. As we know, the voxel values in reconstructed object, linear attenuation coefficient, changes with the energy of photons. However, in conventional CT, we use energy-integrating detectors (EID), which means the reconstructed attenuation map can only reflect a relative intensity (Lell and Kachelrieß 2020). To obtain material specific information, people extend single energy to multiple energies and get convincing results in clinical application (Alvarez and Macovski 1976).

Photon-counting detectors is an emerging technology in the current trend of CT. Different from EID, PCD can count the number of photons that exceed the preset thresholds. By setting multiple energy thresholds, we could get counts in more than two energy bins for one scan. It has potential in improving contrast to noise ratio (CNR), spatial resolution, optimizing quantatitive imaging and reducing radiation dose etc (Willemink er al 2018). However, due to some factors such as cross talk and pulse pile-up, the detected spectrum is non-ideal (Taguchi and Iwanczyk 2013).

To solve this problem, many researches focus on building a model to describe these factors and make corrections accordingly. Schlomka *et al* employed synchrotron radiation to calibrate the detector response function in the energy range of 25-60 keV (Schlomka et al 2008). Then, they used calibrated detector response function in the maximum likelihood estimation of material decomposition and showed the feasibility of quantitative K-edge CT imaging. However, it is difficult to get access to synchrotron radiation. In order to make an easier adapted calibration method, Ding *et al* proposed to use X-ray fluorescence (Ding et al 2014). The simulated results are in good agreement with the measurement of PCD. Based on this technique, Wu *et al* proposed a hybrid Monte Carlo model for detector response function (Wu et al 2016). They used GEANT4 to model the process of X-ray energy deposition in the detector and also used X-ray fluorescence to calibrate the parameters of cross talk model. In addition to X-ray fluorescence, there is also another calibration method using radioisotopes (Taguchi et al 2011). However, the flux rate of both X-ray fluorescence and radioisotopes are relatively lower than X-ray source, which means these calibration methods can't account for pulse pile-up. For pulse pile-up, Wang *et al* analyzed three kind of detector count statistics (delta pulse, unipolar pulse and bipolar pulse) and compared the performance of them (Wang et al 2011). Considering multiple photons pile-up, Taguchi *et al* made a series of research and corrected pulse pile-up for paralyzable detectors and nonparalyzable detectors (Taguchi et al 2010). They also proposed a cascaded method to correct cross talk and pulse pile-up respectively (Cammin et al 2014).

Using the calibrated model, some people tried to directly make basis component decomposition (Michel et al 2009) and others focused on monochromatic imaging by recovering correct counts in each energy bins. Michel *et al* used calibrated detector response function in deconvolution methods to recover incident spectrum from measured data. The width of energy bins in the experiment changed from 0.5keV to 1.5keV, and in this situation, this method showed good performance (Michel et al 2009). Sievers *et al* applied Bayesian deconvolution method to improve the stablility of deconvolved spectrum with the width of energy bins up to 4keV (Sievers et al 2012). Targeting on wider energy bins, Wang *et al* adopted EM algorithm to deconvolve monochromatic spectrum iteratively. They extended the deconvolution from spectrometry to CT reconstruction furtherly and get a satisfying result (Wang et al 2017). Although these methods shows nice results, they depend on the preciseness of the calibrated detector response function and need to scan in a large amount of energy bins, which is impractical in clinical diagnosis.

Recently, deep learning-based methods have shown its great potential in medical imaging such as detection, segmentation and reconstruction (Lell and Kachelrieß 2020). For monochromatic imaging, growing attention has been paid to deep learning type methods. Touch *et al* utilized a two-hidden-layers fully connected neural network to learn to undo the distortion of spectrum in spectral CT. To reduce artefacts and further improve the performance, post-reconstruction bilateral filtration-based denoising method was applied. This method was evaluated in both full-spectrum and four energy bins cases. Both cases showed an improved performance, but 4 bins mode is not so satisfying (Touch et al 2016). Zimmerman *et al* also used a simple neural network to make an experimental investigation of this kind of method. Different from Touch, the output of the neural network is directly set as basis material component in sinogram. They also proposed some transfer learning strategies such as from simulation data to experimental data and from aggregated pixels to individual pixels. And their method got a good performance in K-edge imaging (Zimmerman et al 2020). In addition to this kind of sinogram domain deep learning-based monochromatic imaging methods, there are also some methods working in image domain. Feng *et al* proposed a fully connected neural network for monochromatic imaging. This method used image patch as input and get a robust result in experiment (Feng et al 2018). Shi *et al* adopted a different network, Wasserstein Generative Adversarial Network with a Hybrid Loss and also obtained a good performance (Shi et al 2019).

In real-world scanning, considering radiation dose and noise level, we could only get counts in a certain number of wide energy bins (no more than four). In this situation, the monochromatic imaging problem is ill-conditioned. Conventional deconvolution method can't recover the correct spectrum accurately. Besides, the results of conventional method also affected by the accuracy of calibrated detector response function. For deep learning-based method, compared with image domain monochromatic imaging method, sinogram domain method has a clearer explanation, to undo the distortion and recover correct counts in each energy bins. However, most deep learning-based monochromatic imaging methods just simply combined the basic layer such as fully connected layer, convolution layer and activation layer etc. These methods didn't analysis the problem in detail and consider whether these existing network architectures could correct the spectrum distortion in an efficient way.

Based on these considerations, in this paper, we proposed a deep-learning based monochromatic image synthesis method working in sinogram domain. Focusing on the physical model of cross talk, we design a novel network architecture which is more appropriate for this problem. Our method doesn't need calibrated detector response function and can use data from less amount of wider energy bins. On our CBCT system with a PCD, we made a series of experimental study to demonstrate its feasibility and get very encouraging results. At the end of this paper, we also discuss the potential of our method on pulse pile-up correction.

The rest part of this paper is organized as follows. In section 2, we present the principle of our method. The details and results of experimental study are showed in section 3. In section 4, we make a series of discussions about our method and its potential application. And finally, we get conclusions.

## 2. Methods

To get monochromatic images, we have to make corrections to the distorted spectrum. First, we will model the PCD-CT system.

### 2.1 Problem modeling

Using detector respone function to model cross talk, the whole imaging process of PCD-CT system can be formulated as:

$$\lambda_i = \int_{E_{\text{LT}_i}}^{E_{\text{HT}_i}} dE' \int_0^\infty S_0(E) e^{-\int \mu(E,x)dx} h(E'; E) dE \tag{1}$$

where $\lambda_i$ represents the counts of photon in the $i$-th energy bin ($i = 1, 2, ...$), $E_{\text{HT}_i}$ and $E_{\text{LT}_i}$ are high and low thresholds of this energy bin respectively. $S_0(E)$ denotes the spectrum of X-ray source and $\mu(E, x)$ is linear attenuation coefficient with $x$ being spatial coordinates. $h(E'; E)$ is the detector response function and it represents the probability that a photon of energy $E$ is recorded as $E'$ by the detector. Let $S_{\text{in}}(E) = S_0(E) e^{-\int \mu(E,x)dx}$ and $H(i; E) = \int_{E_{\text{LT}_i}}^{E_{\text{HT}_i}} h(E'; E) dE'$, we have:

$$\lambda_i = \int_0^\infty S_{\text{in}}(E) H(i; E) dE \tag{2}$$

In real system, it is discretized into:

$$\lambda_i = \sum_j N_j H_{ij} \tag{3}$$

where $N_j = \int_{E_L}^{E_H} S_{\text{in}}(E)$ and $H_{ij} = \frac{\int_{E_L}^{E_H} S_{\text{in}}(E) H(i; E) dE}{\int_{E_L}^{E_H} S_{\text{in}}(E)} = \int_{E_L}^{E_H} S_n(E) H(i; E) dE$

It means $H_{ij}$ is not a constant but a weighted average according to the normalized incident spectrum $S_n(E)$. Considering noise and writing equation (3) in a matrix form, we get:

$$\boldsymbol{\lambda} = \mathbf{HN} + \mathbf{W} \tag{4}$$

suppose $\mathbf{W}$ is a Gaussian noise term with zero mean and the covariance matrix is denoted as $\mathbf{R}$. The maximum likelihood estimation of incident spectrum is:

$$\widehat{\mathbf{N}} = (\mathbf{H}^T \mathbf{R}^{-1} \mathbf{H})^{-1} \mathbf{H}^T \mathbf{R}^{-1} \cdot \boldsymbol{\lambda} \tag{5}$$

In medical imaging, the normalized incident spectrum roughly shows a decreasing trend with photon energy and the shape of it will not change too much. Therefore, we denote the mean of $\mathbf{H}$ is $\mathbf{H_0}$ and the corresponding normalized incident spectrum and detected spectrum are $\bar{S}_n(E)$ and $\bar{\boldsymbol{\lambda}}$. Then, we have:

$$H_{ij} = \int_{E_L}^{E_H} \bar{S}_n(E) H(i; E) dE + H_{ij}(\Delta S_n)$$

$$\mathbf{H} = \mathbf{H_0} + \mathbf{H}(\Delta S_n) = \mathbf{H_0} + \mathbf{H}(\Delta \mathbf{N}) = \mathbf{H_0} + \mathbf{H}(\Delta \boldsymbol{\lambda}) \tag{6}$$

The first term $\mathbf{H_0}$ is a constant and the second term $\mathbf{H}(\Delta \boldsymbol{\lambda})$ is a variable. We assume $\Delta \boldsymbol{\lambda}$ is a relatively small term. Combining equation (5) and (6), we have:

$$\widehat{\mathbf{N}} = \left( \left(\mathbf{H_0} + \mathbf{H}(\Delta \boldsymbol{\lambda})\right)^T \mathbf{R}^{-1} \left(\mathbf{H_0} + \mathbf{H}(\Delta \boldsymbol{\lambda})\right) \right)^{-1} \left(\mathbf{H_0} + \mathbf{H}(\Delta \boldsymbol{\lambda})\right)^T \mathbf{R}^{-1} \cdot \boldsymbol{\lambda} \tag{7}$$

In practice, we let the number of detector energy bins equaling to the number of monochromatic energies, which means $\mathbf{H}$ is a square matrix. Because the different energy bins in detector response function is independent, $\mathbf{H}$ is full rank. Therefore, $\mathbf{H}$ is an invertible matrix. And equation (7) turns into:

$$\widehat{\mathbf{N}} = \left(\mathbf{H}_0 + \mathbf{H}(\Delta\boldsymbol{\lambda})\right)^{-1} \cdot \boldsymbol{\lambda} \tag{8}$$

Because $\Delta\boldsymbol{\lambda}$ is a relatively small term, we can do Taylor expansion to $\left(\mathbf{H}_0 + \mathbf{H}(\Delta\boldsymbol{\lambda})\right)^{-1}$ and omit high order terms.

$$\left(\mathbf{H}_0 + \mathbf{H}(\Delta\boldsymbol{\lambda})\right)^{-1} = \mathbf{H}_0^{-1} - \mathbf{H}_0^{-1} \cdot \begin{pmatrix} \left(\frac{dH_{11}}{d\boldsymbol{\lambda}}(\bar{\boldsymbol{\lambda}})\right)^T \cdot \Delta\boldsymbol{\lambda} & \cdots & \left(\frac{dH_{1n}}{d\boldsymbol{\lambda}}(\bar{\boldsymbol{\lambda}})\right)^T \cdot \Delta\boldsymbol{\lambda} \\ \vdots & \ddots & \vdots \\ \left(\frac{dH_{n1}}{d\boldsymbol{\lambda}}(\bar{\boldsymbol{\lambda}})\right)^T \cdot \Delta\boldsymbol{\lambda} & \cdots & \left(\frac{dH_{nn}}{d\boldsymbol{\lambda}}(\bar{\boldsymbol{\lambda}})\right)^T \cdot \Delta\boldsymbol{\lambda} \end{pmatrix} \cdot \mathbf{H}_0^{-1} \tag{9}$$

We denote the matrix between $\mathbf{H}_0^{-1}$ as $\boldsymbol{\Phi}(\boldsymbol{\lambda})$, then we have:

$$\widehat{\mathbf{N}} = (\mathbf{H}_0^{-1} - \mathbf{H}_0^{-1} \cdot \boldsymbol{\Phi}(\boldsymbol{\lambda}) \cdot \mathbf{H}_0^{-1}) \cdot \boldsymbol{\lambda}$$

$$= \left(\mathbf{I} - \mathbf{H}_0^{-1} \cdot \boldsymbol{\Phi}(\boldsymbol{\lambda})\right) \cdot \mathbf{H}_0^{-1} \cdot \boldsymbol{\lambda} \tag{10}$$

Therefore, our object is to use deep learning methods to find a correction function $f_\theta$ with its input being $\boldsymbol{\lambda}$ and output being $\mathbf{N}$.

*2.2 Network architecture*

From equation (10), we know that the correction function $f_\theta$ depends on the detected spectrum $\boldsymbol{\lambda}$, which means for different input $\boldsymbol{\lambda}$, the parameter of correction function $\boldsymbol{\theta}$ is also different ($\boldsymbol{\theta} = \boldsymbol{\theta}(\boldsymbol{\lambda})$). However, for conventional neural network, it is difficult to model this kind of dynamic process efficiently, because the parameter such as convolution kernels are fixed after training. Therefore, we propose a novel dynamic network architecture to solve this problem.

The key idea is to construct a dynamic neural network whose parameters can change with input. To achieve this, some people used attention block to learn an array and weighted the static convolution kernels (Yang et al 2019). Different from this idea, we directly use one part of neural network to output the dynamic parameter and design our network according to equation (10).

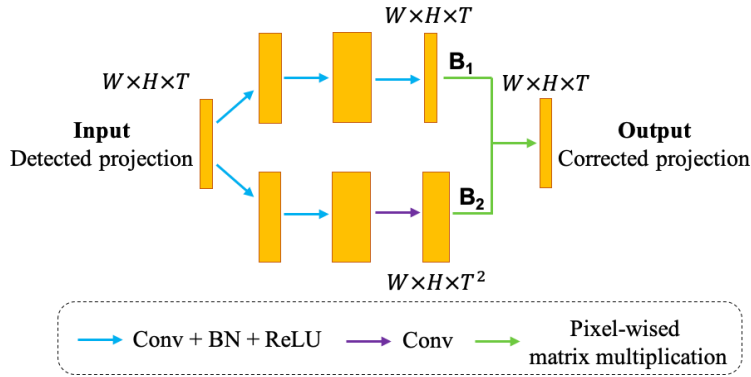

Figure 1. The architecture of neural network

The detailed architecture of neural network is shown in figure 1. We use the detected signal in one view as input with its size being $W \times H \times T$. $W$ and $H$ denote the width and height of the detector panel, and $T$ means the number of energy bins. The branch $\mathbf{B_1}$ learns static part $\mathbf{H}_0^{-1} \cdot \boldsymbol{\lambda}$, and branch $\mathbf{B_2}$ learns to output the dynamic parameters, $\mathbf{I} - \mathbf{H}_0^{-1} \cdot \boldsymbol{\Phi}(\boldsymbol{\lambda})$. These two branches are combined in the green arrow which means pixel-wised matrix multiplication. More concretely, the output of $\mathbf{B_2}$ has a size of $W \times H \times T^2$. We first reshape it to $W \times H \times (T \times T)$, and then, for each pixel in $W \times H$, we do a matrix multiplication with $\mathbf{B_1}$. To make our results less noisy, we could use log-transformed projections for both input and output, and thus add activation layers accordingly to learn the relationship of post-log projections. Finally, to make the network easier to learn, we could also let it learn in a residual way.

Compared with conventional neural network which is a simple stack of convolutional layers and fully connected layers, our architecture is more efficient. It can achieve better results with the same number of parameters, because it can match the physical model of spectrum correction better.

## 3. Experimental study

To illustrate the effectiveness of our method, we made an experimental study on our CBCT system which is shown in figure 2. The PCD is XCounter Flite X1 and the size of it is (1536×0.1mm)×(128×0.1mm). In our experiment, we set three energy bins, which are [27.1, 39.7], [39.7, 50.5] and [50.5, 85] keV respectively and the corresponding monochromatic energy is 35, 45 and 60 keV. To avoid pulse pile-up and illustrate the denoising ability of neural network, we set X-ray tube working at 85kVp and 0.25mA. The total number of photons is about 10000. We make 31 phantoms as our dataset which contains solution such as NaCl (aq), KCl (aq), $CuSO_4$ (aq) with different concentration and PMMA, PET, POM etc. We divided them into 24, 4, 3 for training, validation and test.

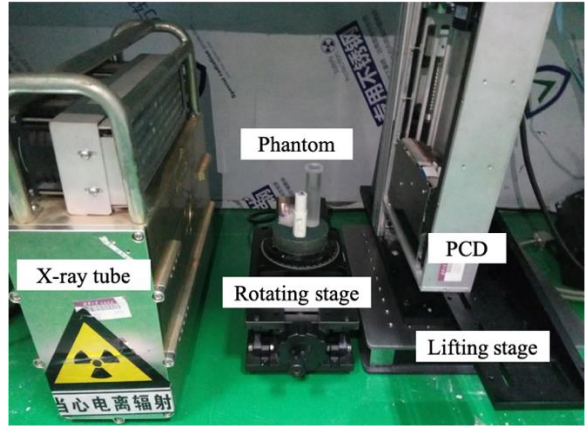

Figure 2. The illustration of our CBCT system.

The input of our network is 2D detected signal in one view and the output is the corresponding corrected monochromatic sinogram. To get label for training, we first directly reconstruct the detected sinogram. Then we made segmentation according to different thresholds to separate different materials. Finally, we use standard linear attenuation coefficient from NIST database to fill in each region and make projection for each monochromatic energy. Considering the relevance between pixel, we set kernel size as 3×3 in our method. The network was trained in a supervised way with loss function being L1 distance and the objective function is shown in equation (11). The optimizer was Adam and learning rate was set as 1e-4. We stopped training when validation loss did not decrease in successive 20 epochs. To evaluate our method in an intuitive way, we use FDK algorithm to reconstruct the output sinogram.

$$\boldsymbol{\theta}^* = \underset{\boldsymbol{\theta}}{\mathrm{argmin}} \sum \left\| \int \boldsymbol{\mu} dx - f_{\boldsymbol{\theta}}(\boldsymbol{\lambda}) \right\| \tag{11}$$

To demonstrate the effectiveness of our method, we compared our method with some other monochromatic imaging methods. For conventional method, we used pre-reconstruction method such as maximum likelihood estimation to make material decomposition in sinogram, and then synthesize monochromatic images (denoted as MD in sinogram). The detector response function was calibrated according to Wu's method (Wu et al 2016) and we chose Al and PMMA as basic materials. For neural network method, we used touch's method (Touch et al 2016) for comparison (denoted as FC Net). The input and output are same with our method. But he used a two-hidden-layers network to learn spectrum correction.

*3.2.1 linear attenuation coefficient*

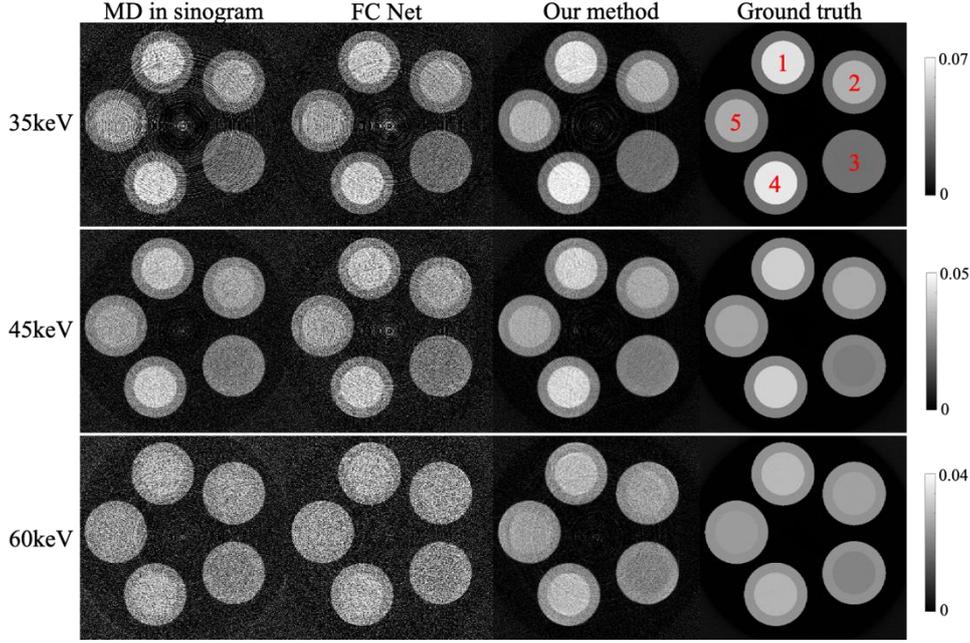
Figure 3. Reconstruction images from different methods.

Table 1. Quantitative comparison of linear attenuation coefficients from different methods.

| ROI | Materials | MD in sinogram | FC Net | Our method | Ground truth |
|---|---|---|---|---|---|
| 1 | 25% NaCl (aq) | (35 keV) 0.0562<br>(45 keV) 0.0370<br>(60 keV) 0.0268 | 0.0523<br>0.0366<br>0.0276 | **0.0612**<br>**0.0403**<br>**0.0283** | 0.0620<br>0.0405<br>0.0289 |
| 2 | 15% NaCl (aq) | 0.0441<br>0.0308<br>0.0236 | 0.0424<br>0.0309<br>0.0244 | **0.0475**<br>**0.0328**<br>**0.0249** | 0.0484<br>0.0335<br>0.0253 |
| 3 | $H_2O$ | 0.0277<br>0.0226<br>0.0196 | 0.0281<br>0.0221<br>**0.0200** | **0.0303**<br>**0.0238**<br>0.0198 | 0.0307<br>0.0243<br>0.0205 |
| 4 | 10% $CuSO_4$ (aq) | 0.0598<br>0.0385<br>0.0272 | 0.0551<br>0.0382<br>**0.0277** | **0.0656**<br>**0.0418**<br>0.0288 | 0.0636<br>0.0407<br>0.0281 |
| 5 | 5% $CuSO_4$ (aq) | 0.0436<br>0.0305<br>0.0233 | 0.0420<br>0.0305<br>0.0238 | **0.0476**<br>**0.0326**<br>**0.0244** | 0.0471<br>0.0325<br>0.0243 |
| Average relative error | | (8.27±1.93)%<br>(7.00±0.77)%<br>(5.32±1.11)% | (12.11±1.60)%<br>(7.80±1.20)%<br>(3.01±1.33)% | (1.73±0.62)%<br>(1.62±0.43)%<br>(1.98±0.46)% | |

The corresponding results of linear attenuation coefficient are showed in figure 3 and table 1. Compared with other methods, our method showed better performance on both visual and quantitative evaluation. MD in sinogram method depends on the preciseness of X-ray spectrum and calibrated detector response function. In practice, it will also be affected by the stability of X-ray source. According to our result and touch's paper, FC Net could get relatively good results in full spectrum mode. However, we can't get projection data in so many energy bins in clinical usage. In three energy bins situation, FC Net can't solve this ill-posed problem well because of its small amount of parameters and simple structure. But our method can get more precise monochromatic images with high quality.

*3.2.2 Basic material coefficient*

One of the important applications of spectral CT is material decomposition. Therefore, in addition to linear attenuation coefficient, we used post-reconstruction material decomposition method to further demonstrate the potential of our method.

We also chose Al and PMMA as our basic materials. Using the linear attenuation coefficient in different energies through our method, our object is to solve $\alpha_{Al}$ and $\alpha_{PMMA}$ in equation (12).

$$\mu(E) = \alpha_{Al}\mu_{Al}(E) + \alpha_{PMMA}\mu_{PMMA}(E) \qquad (12)$$

The decomposition results are shown in figure 4 and table 2.

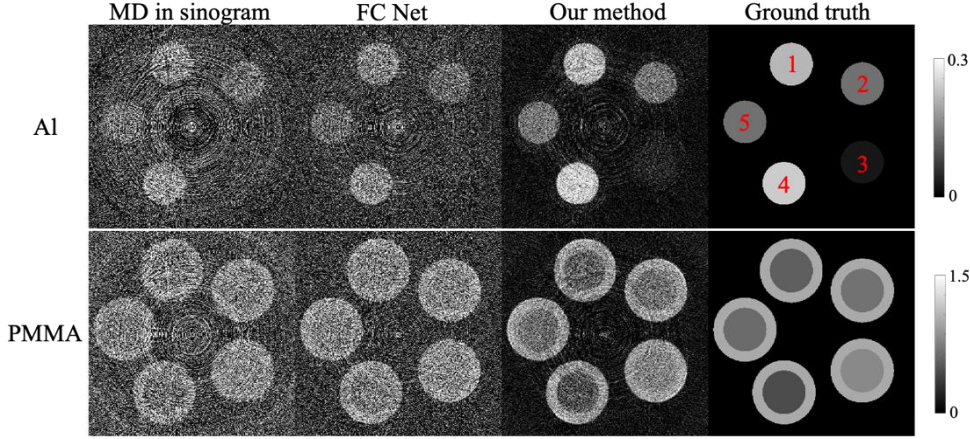

Figure 4. The results of material decomposition.

Table 2. Quantitative comparison of basic material coefficient.

| ROI | Materials | MD in sinogram | FC Net | Our method | Ground truth |
|---|---|---|---|---|---|
| 1 | 25% NaCl (aq) | (Al) 0.1820 (PMMA) 0.5738 | 0.1402 0.7501 | **0.2135** **0.5465** | 0.2154 0.5579 |
| 2 | 15% NaCl (aq) | 0.1084 0.6762 | 0.0888 0.7754 | **0.1311** **0.6553** | 0.1329 0.6707 |
| 3 | $H_2O$ | 0.0054 0.8369 | 0.0135 0.8090 | **0.0324** **0.7604** | 0.0274 0.8072 |
| 4 | 10% $CuSO_4$ (aq) | 0.2118 0.4934 | 0.1637 0.6889 | **0.2496** **0.4459** | 0.2399 0.4471 |
| 5 | 5% $CuSO_4$ (aq) | 0.1081 0.6635 | 0.0924 0.7379 | **0.1380** **0.6128** | 0.1330 0.6300 |

From the above results, we know that our method showed better performance not only on linear attenuation coefficient but also on basic material coefficient. The average relative errors of our method over the five ROIs are 5.69% for Al coefficient and 2.63% for PMMA coefficient. And this demonstrated our method has good potential on material differentiation. We also notice that the noise of figure 6 is heavy. This is because the number of photons in our experiments is quite small, which is about 10000. And compared with linear attenuation coefficient, the result of material decomposition is more sensitive to noise.

## 4. Discussion and conclusion

For monochromatic imaging, the key problem is to corrected spectrum distortion. This problem will become ill-posed when we could only get counts from a small number of wide energy bins. In section 2, considering detector response function in this situation, we analysed the problem model and found it is a dynamic process. It means some parameter of distortion model is dependent on incident spectrum. Therefore, the correction should also be dynamic. Based on these analysis, we proposed a novel network architecture to correct the spectrum. Diffferent from conventional neural network, by changing the way of basic operation, we make some parameters of our network change from static to dynamic. In this way, our network parameters will also change with different input. From the analysis of section 2, it is obvious that our network architeture is more suitable. Moreover, there is also another viewpoint to explain our method. By changing basic operation rule of network, we increased the complexity of network with same number of parameters. And this will make the network more efficient. About dynamic neural network, we also found some research in computer vision (Chen et al 2020), which had some similarities with our idea but more differences in motivation and implementation.

Compared with other monochromatic imaging methods, our method showed good potential on precise monochromatic reconstruction, which make it feasible for quantitative imaging. For other applications such as material separation (Rigie and La Rivière 2016) and kidney stone calssification (Marcus et al 2018, Weisenthal et al 2018), we may focus more on other information. Therefore, in addition to linear attenuation coefficient, we also used basic material coefficient to evaluate our method. And our method also performed well. Because the dose of our experiment is relatively low (10000 photons), conventional methods showed severe artefacts especially for material decomposition. By comparison, our method showed a good ability of denoising.

Besides the above experiments, there are still some further work to be done. First, targeting on pulse pile-up, we want to find out how our method would perform on data scanning with big tube current. Second, for different application, we need to make our method more specific. This means we should optimize the choice of erergy bins and the category or shape of our datasets. Finally, our method shall be tested on different PCD.

In this paper, we proposed a deep learning-based method for monochromatic imaging with limited number of energy bins. Different from conventional methods, our method can avoid the unpreciseness of detector response function and learn to correct detected spectrum from data. By analyzing the model of spectrum distortion, we designed a novel network architecture with dynamic parameters. Through experimental study, our method performed better than other monochromiatic imaging methods on not only linear atttenuation coefficient but also basic material coefficient. This result showed encouraging potential of our method on quantitative imaging and material separation.